# Electrosense+:
# Crowdsourcing Radio Spectrum Decoding using IoT Receivers

Roberto Calvo-Palomino[a], Héctor Cordobés[a], Markus Engel[c], Markus Fuchs[c], Pratiksha Jain[d], Marc Liechti[d], Sreeraj Rajendran[b], Matthias Schäfer[c], Bertold Van den Bergh[b], Sofie Pollin[b], Domenico Giustiniano[a], Vincent Lenders[e]

[a]*IMDEA Networks Institute, Madrid, Spain*
[b]*Department ESAT, KU Leuven, Belgium*
[c]*SeRo Systems, Germany*
[d]*Trivo Systems, Bern, Switzerland*
[e]*armasuisse, Thun, Switzerland*

**Abstract**

Web spectrum monitoring systems based on crowdsourcing have recently gained popularity. These systems are however limited to applications of interest for governamental organizations or telecom providers, and only provide aggregated information about spectrum statistics. The result is that there is a lack of interest for layman users to participate, which limits its widespread deployment. We present Electrosense+ which addresses this challenge and creates a general-purpose and open platform for spectrum monitoring using low-cost, embedded, and software-defined spectrum IoT sensors. Electrosense+ allows users to remotely decode specific parts of the radio spectrum. It builds on the centralized architecture of its predecessor, Electrosense, for controlling and monitoring the spectrum IoT sensors, but implements a real-time and peer-to-peer communication system for scalable spectrum data decoding. We propose different mechanisms to incentivize the participation of users for deploying new sensors and keep them operational in the Electrosense network. As a reward for the user, we propose an incentive accounting system based on virtual tokens to encourage the participants to host IoT sensors. We present the new Electrosense+ system architecture and evaluate its performance at decoding various wireless signals, including FM radio, AM radio, ADS-B, AIS, LTE, and ACARS.

## 1. Introduction

The idea of web-based distributed radio applications has recently gained interest such as WebSDR[1], LiveATC[2] and ElectroSense [1], motivated by the diversity in space of the spectrum and the wide range of services benefiting from it. Multiple crowdsourcing initiatives have been proposed using various spectrum sensors ranging from low-end hardware [2–6] to expensive spectrum analysers [7, 8]. These initiatives monitor the electro-magnetic spectrum in a distributed way and provide applications that target specific communities.

---

[1]http://www.websdr.org/
[2]https://www.liveatc.net/

Some of the major initiatives for analyzing the entire wireless electromagnetic spectrum are Electrosense [1], Microsoft Spectrum Observatory [9], Google TV White Space [10], IBM Horizon [11] and SpecNet [3]. Other initiatives focus instead on more specific monitoring applications over a limited frequency range, such as remote radio monitoring stations in OpenWebRX [12], KiwiSDR [13] and WebSDR, live air traffic control (ATC) broadcasts from air traffic control towers in LiveATC or aircraft monitoring systems such as OpenSky [14]. Airspy [15] provides a sensor client-server architecture to operate SDRs remotely, but it relies on the computational power of the client-side to decode the signals, and high network bandwidth to send I/Q data stream to the client.

All the above initiatives have severe drawbacks, such as limited use cases (e.g., focus only on FM radio decoding or spectrum analysis), lack of interest for layman users to host a sensor (dynamic spectrum access and anomaly detection do not attract the large audience), require expensive SDRs or dedicated hardware (such as the Microsoft Observatory), poor scalability and complicated process to run measurement campaigns or access the data (sensors are busy), or high network requirements for sending I/Q data to the client.

Our vision is that people are the primary operators of spectrum sensors. We aim at empowering people implementing a global spectrum monitoring system which let them connect to any spectrum sensor in the network and decode any publicly decodable radio spectrum part, such as broadcast and control messages, in real time through the Internet. In our system, spectrum analysis, or applications such as dynamic spectrum access and anomaly detection become secondary tasks, being active only if the sensor is not used by people. The overarching goal is to support low-cost and software-defined IoT (Internet-of-Things) [16] spectrum sensing devices and to provide incentives for people to participate and host those sensors at their homes or organizations, enhancing the mission of building a crowdsourcing spectrum monitoring system.

Our contributions are:

- We propose a novel radio spectrum decoding architecture where the primary operators are the users. The architecture provides a transparent system to decoding the spectrum on the embedded sensors, and makes use of real-time peer-to-peer communication to send the information already decoded to the users.

- We implement the decoding process on the spectrum sensors in an efficient way alleviating the processing load in the client, reducing the network bandwidth used, and adding a security-privacy layer since no raw data (I/Q) is sent to the users.

- We introduce an incentive for sensors' owners to be part of the radio crowdsourcing community based on tokens. We propose a user rewarding system which also regulates the sensor usage rights in a fair manner for all users.

- We evaluate the architecture proposed in real scenarios with 6 different decoders: FM/AM radio, ADS-B, AIS, LTE, and ACARS. We compare our solution proposed in this work with the existing related projects.

- We release Electrosense+ website publicly. Sensing module executed on the IoT sensor and the API[3] are released as open source to facilitate the integration of future decoders[4].

---

[3] https://electrosense.org/api-spec
[4] https://github.com/electrosense/es-sensor

## 2. Design Goals

Past web-based spectrum monitoring initiatives are either application-specific [12, 17] or do not scale well for remote signal decoding [1]. Scalability is challenged by the large data volumes needed for wideband spectrum monitoring, much higher than needed by typical IoT applications. But, even with a larger bottleneck, we experienced that motivating users to deploy sensors and keep their sensors operational is the main hurdle for the wide-spread deployment of crowdsourced spectrum monitoring. The main reason is that the most interesting services for stakeholders that need to monitor the spectrum, such as governmental organizations and telecom providers, are orthogonal to the interests of the vast majority of users.

In this work, we propose a novel radio monitoring architecture that addresses the main limitations of previous systems:

***General purpose decoding.*** The system architecture allows to decode any public decodable wireless technologies that is within range of the deployed sensors. As the spectrum is used by many different wireless technologies and new technologies are emerging constantly, we support the integration of open source spectrum decoders developed by the community. The system architecture thus defines open interfaces and APIs to allow easy integration of various decoder types.

***Peer-to-peer architecture.*** As the system is expected to support a large number of concurrent users and spectrum data is very large in nature, a centralized approach is unfeasible as it would cause a data deluge to the backend or large latency from the sensor to the consumer. In order to support real-time applications and scale well, Electrosense+ supports peer-to-peer communication between the spectrum sensors and the users.

***Sensor owners incentive.*** Since in crowdsourcing initiatives, people are expected to acquire and run a spectrum sensor node on their own, good incentives are needed to foster participation. This includes rewards for hosting a spectrum sensor but also to provide valuable spectrum services that they will get in return. In Electrosense+, spectrum services are provided to the users in the form of *spectrum apps* and sensor owners receive tokens for the time their sensors are online and used by the community.

***Security & privacy.*** Spectrum data can contain private information and there should be limitations on some specific frequencies about the information type that users can listen to. For example, the system should not allow users to listen to private voice or other text conversations. It can instead decode broadcast and control messages. To this end, Electrosense+ does not transmit raw I/Q data to the users but only *aggregated* spectrum data and *filtered* decoded data. That way, Electrosense+ keeps full control over the data that users will receive by enforcing strict integration policies on which decoders are allowed to run on the sensors and data is filtered.

## 3. Architecture

The Electrosense+ architecture is depicted in Figure 1. The main system components are the sensor, the client, and the backend. While these components were all present in the original Electrosense design, the novelty is to enable direct peer-to-peer connections between the sensor and clients in order to provide direct decoding services (apps) for users, and to account for the usage patterns in order to reward sensor operators. In this section, we focus on the new required architectural components for these enhancements while we refer to [1] for a description of the original Electrosense design. Electrosense+ is fully backward compatible with previous versions of the system.

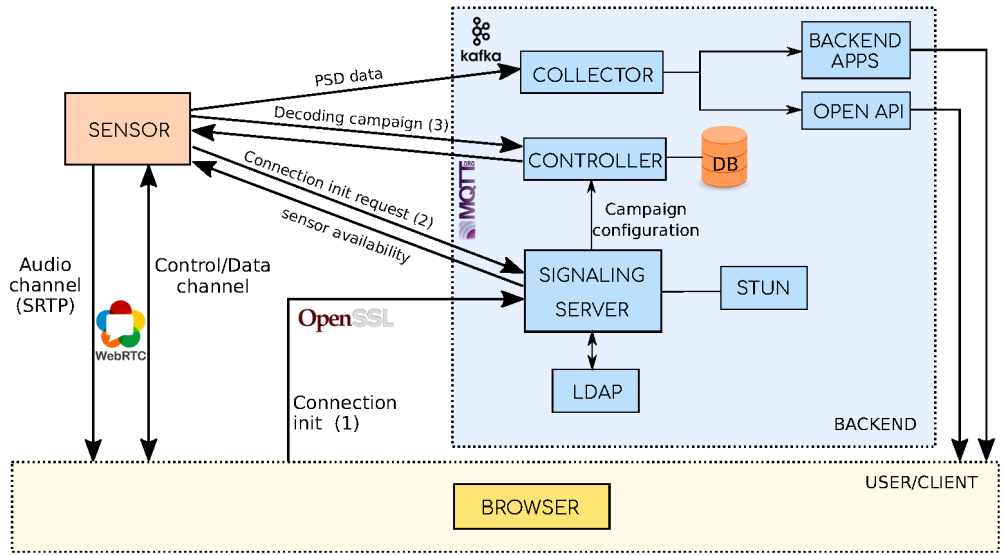

Figure 1: Full overview Electrosense+ architecture

*3.1. Signaling and Controlling*

The sensors are managed by the Electrosense+ backend over secure messaging via the MQTT [18] protocol. The client-sensor connection is handled over the WebRTC protocol suite. WebRTC [19] is used in well-known applications such as Google Hangouts and Facebook Messenger providing real-time communication (RTC) capabilities integrated in web browsers without using dedicate plugins for this task. At an initial communication state, client and sensor need to exchange meta-data to coordinate the communication using the *signaling server*. The Session Traversal Utilities for NAT (*STUN*) server allows to find the public IP address of the client and sensor in order to provide a direct connection between them, even if they are located behind firewalls or Network Address Translators (*NAT*).

When a client wants to connect to a sensor, it signals a request through the signaling server and establishes a direct connection to the sensor (without passing through the Electrosense+ backend). Then, two different channels are created: control/data channel and Audio channel (as Figure 1 shows). The control/data channel is a bi-directional channel used to send the spectrum information and data decoded from the sensor to the client, and to command the sensing parameters from the client to the sensor (frequency, gain, etc.). The Audio channel is exclusively used to stream audio in real time from the sensor to the client using Secure Real Time Protocol (*SRTP*). This peer-to-peer communication minimizes the network delays between the client and sensor, providing a fast and scalable access the data from the sensor node.

When there are no users that are connected to a particular sensor, the sensor is instructed by the controller to sweep the spectrum or launch a specific campaign. As soon as a client connects to a sensor the peer-to-peer communication is established. Then, the client gets access to the sensing parameters (frequency, gain, etc.). Clients can remotely tune to any radio frequency and thus influence which decoder is activated in the sensor, e.g., if the RTL-SDR is tuned to the FM radio band, the FM decoder will be active. If the client tunes to a frequency which has no associated decoder on the sensor node, the client only sees a real-time waterfall diagram of the PSD (*Power Spectral Density*) data at the selected frequency band.

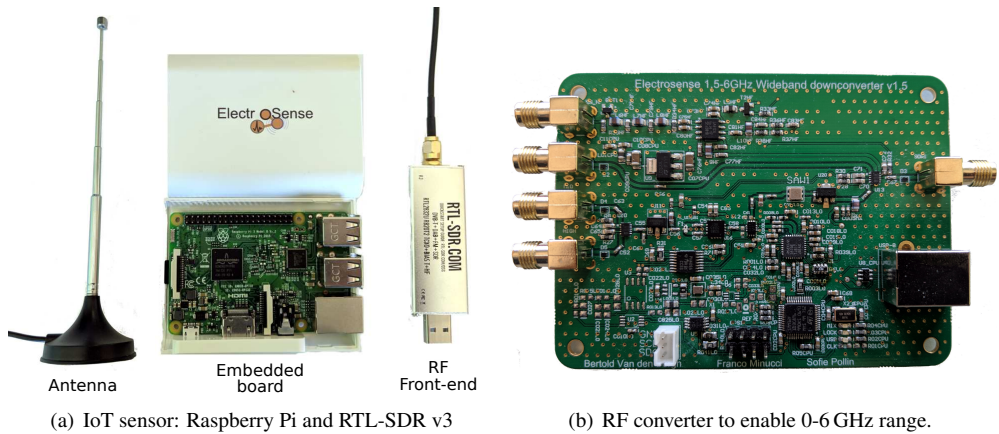

(a) IoT sensor: Raspberry Pi and RTL-SDR v3    (b) RF converter to enable 0-6 GHz range.

Figure 2: Electrosense+ IoT spectrum sensor.

*3.2. Sensor Node*

The software of the sensor node is designed to run on low-cost embedded computing platforms. Figure 2(a) shows the current hardware configuration of the Electrosense sensor which makes use of a Raspberry Pi device for the signal processing and RTL-SDR v3 [20] as radio front-end. The RTL-SDR v3 contains a Temperature Compensated Crystal Oscillator (TCXO) that provides an excellent short-term oscillation frequency stability in changing-temperature environments [21] allowing a better decoding performance. The sensors can measure the RF spectrum ranging from 0 MHz up to 6 GHz using an optional down-converter [1] shown in Figure 2(b).

The Electrosense+ sensor architecture supports two signal processing pipelines in parallel as Figure 3 shows: the *spectrum analysis* (PSD) pipeline and a *decoding* pipeline. Both pipelines are reading the same I/Q data streaming from the RTL-SDR, but they process the data in a different way. The spectrum analysis pipeline computes an aggregated PSD signal representation using the Welch method with implementation based on the Fast Fourier Transform (FFT). PSD data are then sent both to the backend and directly to the connected client. In particular:

- The PSD data is sent from IoT spectrum sensors to the backend. The PSD data is stored for historical inspection of the spectrum and to understand the evolution of spectrum activities over time. This PSD data is accessible by every Electrosense's user through the API [1].

- At the connected client, the PSD data is useful for the user to visually analyze the spectrum in the frequency domain in real time, and to identify parts of the spectrum with ongoing transmissions. Although visualization of PSD data through the backend is also possible, direct connection from the IoT sensor to the client allows for smaller latencies.

The decoding pipeline is used to locally demodulate and decode the signals at the sensor. We implement data decoding in the sensor node as it largely reduces the amount of data sent to the user. In addition, it avoids security and privacy concerns as no I/Q data is sent directly to the users which would allow them to decode any wireless signals, even those that may contain sensitive personal data. The sensor has multiple decoders to decode different parts of the spectrum. So far, we have integrated existing open source decoders including FM radio, AM radio, ADS-B / ACARS (air traffic signals), AIS (ships tracking), and LTE cell broadcasting. In the future, we plan to integrate many more technologies such as e.g., DAB, DVB-T, GSM cell broadcasting, LORA, Sigfox, etc. Given the limited resources of the embedded computing sensor platform, only one decoder is active on each sensor at the same time.

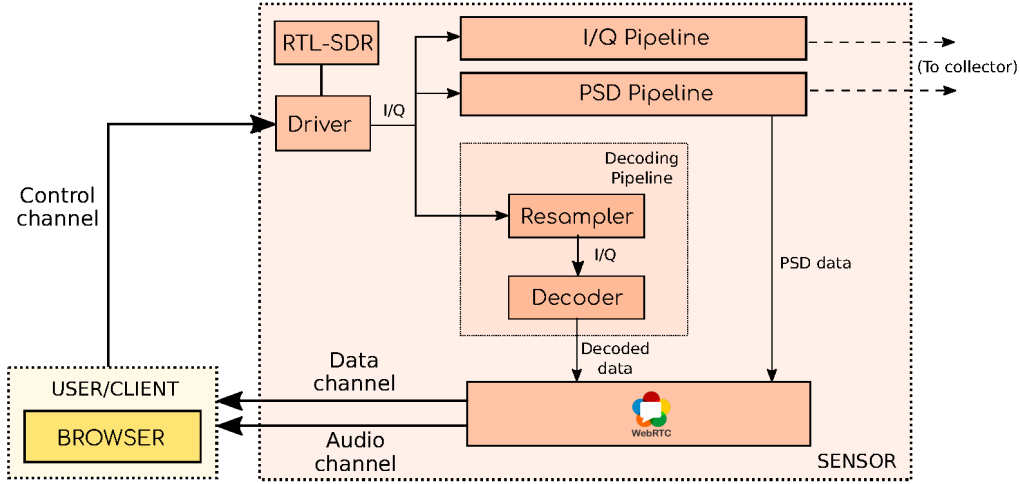

Figure 3: User-Sensor communication diagram.

We have defined standard input and output interfaces to facilitate the integration of existing and future decoders that may be provided by the open source community. The input interface (used between the sensing software and the decoder) is an UDP socket over which the decoder can read the sampling parameters and I/Q data in chunks. For the output interfaces (used between the decoder and web-browser), we differentiate between a message and a streaming interface depending on the type of data that is decoded. Text data is sent as a JSON object over an UDP socket. For audio and video, the data is streamed over an UDP socket in a raw format.

Since the sampling rate that the decoders expect may be different from the configured RTL-SDR sampling rate, a re-sampler is implemented in front of the decoder to provide the sampling rate that the decoder requires. The re-sampler is also useful to decimate the I/Q data for performance reasons as some decoders consume too much CPU processing power on the limited hardware of the sensors when the sampling rate is too large.

*3.3. Reward Model*

Incentives for motivating people have been studied [22–24] in crowdsourcing systems focusing in the bootstrap stage of the project, where a significant scale of deployment is needed to reach a correct system operation where sensor owners can get benefits.

We propose a virtual accounting system based on virtual tokens, that provides two main features in our system. First, it helps to regulate the access to the sensors and distribute the rights among users in a fairly manner. And second, it is used as an incentive for people to host sensors since they will be rewarded for deploying and hosting sensors.

Table 1: Rewarding system notation

| Notation | Description |
|---|---|
| $K$ | Token of the network. |
| $P_{psd}$ | Tokens paid for sensing time in PSD campaign |
| $P_{dec}$ | Tokens paid for sensing time in DEC campaign |
| $R_{s,psd}$ | Reward obtaining by executing PSD campaign |
| $R_{s,dec}$ | Reward obtaining by executing DEC campaign |
| $D_s$ | Sensor density in a region |
| $T$ | Total live time of the network |
| $t_s$ | absolute time when a sensor is deployed ($\in [1, T]$) |
| $t_o$ | total operational time of a sensor ($\in [0, T]$) |
| $E_o$ | Earnings of sensor owner |
| $C_n$ | Cost of the network |
| $E_n$ | Earnings of the network |
| $B_n$ | Benefits of the network |

The sensor provides two operational modes: PSD and decoding. If no users are consuming data from the decoding pipeline, the sensor will operate in PSD mode. A user spends tokens consuming data from the decoding pipeline of a specific sensor. Then, those tokens will be used to first, reward the sensor owner and second, sustain the network.

**Rewarding model for sensor owners.** Our rewarding model issues tokens to users which operate Electrosense+ sensors. The model rewards people hosting sensors depending on three different factors.

- *Sensor density*: The less density of sensors in a region, the greater the owner's earning. In this way the model incentives people to deploy sensors in areas not currently covered.

- *Deployment time*: The model rewards sensors deployed at the beginning of the project, to motivate people join the initiative at early stage to make the network grow quickly.

- *Operational time*: The longer the sensor is operating in any pipeline, the higher the earnings for the sensor owner.

More formally, the user earnings in a given time *t* is computed as follows (see Table 1 for better understanding):

$$E_o = R_{s,psd} + R_{s,dec} \qquad (1)$$

$$R_{s,i} = P_i \cdot \underbrace{\frac{1}{\log(D_s)}}_{\text{sensor density}} \cdot \underbrace{\frac{T}{t_s}}_{\text{deployment time}} \cdot \underbrace{\frac{1}{1 + e^{-t_o}}}_{\text{operational time}}, \qquad (2)$$

where $i \in \{psd, dec\}$.

$$E_o = R_{s,psd} + F \cdot R_{s,dec}, \tag{3}$$

where $F \in \{0, 1\}$.

Therefore the cost (*C*), earnings (*E*) and benefits (*B*) of the network (*n*) are:

$$C_n = \sum_{s=1}^{\infty} (R_{s,psd} + F \cdot R_{s,dec}) \tag{4}$$

$$E_n = \sum_{s=1}^{\infty} (1 - F) \cdot R_{s,dec} \tag{5}$$

$$B_n = E_n - C_n; \tag{6}$$

In other words, the tokens paid by an user for using the decoding pipeline are split (using factor *F*) between the owner of the sensor and the network. Therefore the profitability of the network depends on 1) how much the decoding pipeline is used and 2) the factor *F*.

To avoid abuse, Electrosense+ backend checks regularly the sensor quality, using the frequency of connections of other users to the sensor as metric. The latter is most likely a good indicator that the sensor is deployed at a good location with a good antenna. This approach could also be combined with more sophisticated algorithms in the backend based on signal learning capabilities [25] and anomaly detection [26].

*3.4. Security and Privacy*

While the Electrosense+ architecture allows in principle to decode any type of wireless signals that fall in the frequency range of our sensors, we enforce a strict policy on the allowed decoders in the sensors in order to prevent from disclosing personal information to Electrosense users. The decoders and their operational frequencies are set by Electrosense+ to make sure that any decoded data provided by sensors does not violate any private information. Although the user can propose, implement and even run new decoders on the sensor side, the backend architecture and the web interface will not allow untrusted decoders avoiding privacy leaks. Furthermore, dedicated GUI interfaces are implemented in the Electrosense+ network only for trusted decoders. Figure 1 shows how the signaling server is the responsible of initiating the connection between the browser (client) and the sensor, as long as the credentials and selected decoder are correct. Therefore, users are not allowed to integrate and deploy new decoders in the Electrosense+ network by themselves.

By allowing users to listen to the content of such communication would violate the privacy of people using those devices. Our policy is thus to integrate only decoders for broadcast communication systems and for public control signalling messages. For example, in this work, we have implemented decoders for FM/AM radio, ADS-B and AIS (broadcasting systems). These decoders are integrated in the decoding pipeline on the sensor (see decoder block in Figure 3) where they read IQ samples as input and send the decoded data to the user through the data channel. The implementation of every decoder depends on the signal to be decoded. For communication systems such as LTE or ACARS, we only decode the signalling and management messages which are sent broadcast over the channel. As these messages are meant to be received by all receivers in the neighborhood, they do not pose a threat to the privacy of the users. Electrosense network does not accept or integrate decoders that aim to decode unicast communication.

Table 2: Decoders, operational frequencies/bandwidths and open source projects.

| Decoder | Frequencies | Bandwidth | Project |
| --- | --- | --- | --- |
| AM radio | 153 kHz - 30 MHz | 60 kHz | SciPy.org |
| FM Radio | 88-108 MHz | 240 kHz | SciPy.org |
| acars | 129-136 MHz | 2.4 MHz | acarsdec |
| AIS | 161-162 MHz | 1.6 MHz | rtl-ais |
| ADS-B | 1090 MHz | 2.0 MHz | dump1090 |
| LTE-Cell | 700 - 3500 MHz | 1.92 MHz | LTE-Cell |

*3.5. Spectrum Applications*

Spectrum services are provided to the users by means of spectrum apps that give valuable information. The Electrosense+ architecture is scalable and versatile enough to empower all use cases. The most widespread use of the wireless spectrum is however to broadcast information, and a primary set of spectrum apps focuses on the decoding of such broadcast information. As this broadcast information is intended for the general public, and not encrypted, hence there are no privacy concerns.

A first generation of Electrosense+ spectrum apps focuses on the decoding of broadcast signals that are of interest to a broad audience, such as AM and FM signals, but also ADS-B, AIS or ACARS messages and LTE cells (see Table 2). These applications make use of the lower frequency bands, making it easier to achieve good coverage with a limited number of Electrosense+ sensors. As the system scales to higher deployment densities, also higher frequency signals can be added. We note that the frequencies for AM signals are not covered by a standard RTL-SDR dongle, but with the Electrosense expansion board [1] also lower (and higher) frequencies ranging from DC to 6 GHz can be covered.

## 4. User Interface

The user interface plays an important role in this work to empower people to use Electrosense+ for decoding radio signals of the electromagnetic spectrum. Using standard technologies that execute in web browsers allows us to reach a great number of users which do not have any signal processing knowledge. Figure 4(a) shows the interface when the user has selected the FM radio decoder and tunes the receiver at 105 MHz. The user can visually inspect the spectrum in that band by checking the PSD data. Although the decoder focuses in a narrow band for decoding the FM radio channel (180 kHz), we show the spectrum information of a wider band (2.4 MHz) for a better understanding of the spectrum by the user. The user can distinguish where transmissions are going on by checking the power in different frequencies. The user can click on the interested channel and he/she will start listening the current FM radio station. This audio streaming is sent using the direct audio channel between the sensor and the client. The user also can set different sensing parameters as the DC gain or the volume used by the decoder, or even the power scale to identify better the spectrum and transmissions. In order to help users to identify where the transmissions are, an automatic channel identification algorithm runs in the browser for detecting the transmissions using the PSD data, and identifies where FM radio channels are located. A red rectangle is shown in the web interface for every FM radio channel detected (see Figure 4(a)) which makes it easier for the user to select another FM radio channel.

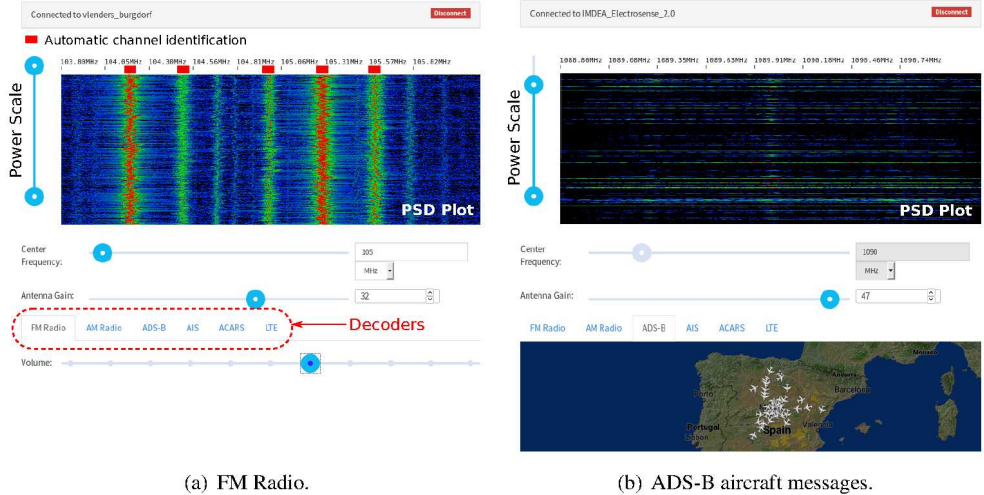

(a) FM Radio.  (b) ADS-B aircraft messages.

Figure 4: User interface for visualizing the spectrum and decoding.

Figure 4(b) shows the interface when the user configures the sensor to decode ADS-B messages (aircraft continuously broadcast their position and other control information). Using the data channel, the sensor sends the PSD data to the client and also sends the information decoded on the sensor. In this case, the ADS-B decoder decodes the messages that aircraft send with information about their location. All this information is collected by the sensor and shown to the user in a map in the web interface.

## 5. Evaluation

In this section, we evaluate the performance of Electrosense+ as a real-time system for decoding the spectrum remotely, the rewarding model and the scalability of the system. The Electrosense+ sensors are based on the RaspberryPi-3B+ and the RTL-SDR radio receiver [20]. The RaspberryPi-3B+ has a Cortex-A53 processor and 1 GB of RAM, representing a typical low-cost IoT device.

### 5.1. CPU Load and Throughput

The software executed on the RaspberryPi-3B+ is split in three main components: (1) sensing, which is responsible to execute the different pipelines to provide PSD data to the client and IQ data to the decoders; (2) WebRTC, which manages the communication between the sensor and the client; and (3) signal decoders. Table 3 shows the CPU load for the three software components and for every implemented decoder. The minimum CPU load of the sensing component (11%) occurs when only the PSD data is computed. When one of the decoder is enabled, the sensing process also needs to execute the resampler process, increasing the CPU load (14-24%). The load depends on the bandwidth and sampling rate that is expected by the decoder (see Table 2). The WebRTC CPU load is very low (0.1-0.2%) since the only task is to manage the communication of the data/audio channel. For the case that FM/AM decoders are enabled, the WebRTC component also needs to resample the audio stream to make the audio compatible with the expected audio input of the browsers. This increases the CPU load up to 6% for this component. The CPU load for every decoder and the total CPU is shown in Table 3. The decoders who use the CPU the most are FM radio and LTECell, but still the RaspberryPi has enough resources to properly process and deliver the decoded data in real time. While the FM/AM radio decoders have a constant CPU usage, other decoders such as LTECell only consume CPU resources for a specific amount of time ($\approx 10$ sec.) to compute the information that is delivered to the client.

Table 3: CPU Load on the Electrosense+ sensor and throughput for every decoder.

| Decoder | CPU (%) sensing | CPU (%) Web-RTC | CPU (%) decoder | CPU (%) Total | Throughput (kb/s) |
|---|---|---|---|---|---|
| PSD | 11 | 0.1 | - | 11.1 | 120-140 |
| FM radio | 24 | 6.5 | 24.6 | 55.1 | 40-50 |
| AM radio | 22 | 6.2 | 10 | 38.2 | 40-50 |
| ADS-B | 14 | 0.2 | 4 | 18.2 | 190-200 |
| AIS | 20 | 0.1 | 6.2 | 26.3 | 50-60 |
| acars | 23 | 0.1 | 8.1 | 31.2 | 90-100 |
| LTECell | 21 | 0.2 | 48 | 69.2 | 10 |

The network bandwidth is also shown in Table 3. The data throughput for AM/FM decoders together with PSD is less than 200 kb/s, which is reasonable for most of the broadband internet connections nowadays. Other existing solutions like Airspy [15] (based on rtl_tcp) make a more intensive use of the network reaching 300-1000 kb/s for the same AM/FM decoding purpose. The maximum network throughput is used by the ADS-B decoder which together with the PSD data reaches 350 kb/s, still a data rate affordable for most home Internet users.

### 5.1.1. Comparison with existing solutions

We perform a comparison between Electrosense+ and other existing projects in terms of CPU and network bandwidth used in the user side. All tested solutions allow to decode spectrum remotely and show information already processed in the web-browser. We have selected AM decoder to perform this comparison since is the common one along all the solutions tested. We use Mozilla Firefox web-browser to perform this test. Figure 5 shows the comparison between Electrosense+ and KiwiSDR, WebSDR and OpenWebRX. Electrosense+ performs slightly better than other solutions in both metrics CPU usage and network throughput.

### 5.2. Real Time Response

Since the new Electrosense+ architecture is built to provide a user experience close to real time for decoding signals, it is important to measure the time delays for the main tasks that can be performed by the user in the user interface (UI). We want to measure the response time when the user accesses the sensor for the first time and the time between the moment a user selects a sensor until the first batch of PSD data flows in the web-browser. The average delay across the sensors attached to Electrosense+ for the user to access the sensor for the first time is 0.97 seconds. This includes the time to establish the WebRTC connection between the browser and the sensor (see Figure 1), tuning the radio frontend to the frequency of operation, collect the IQ stream, process the first batch of IQ data through the PSD pipeline (see Figure 3) and deliver the first batch of processed PSD data through the data channel to the client (browser). This includes the time to establish the WebRTC connection between the browser and the sensor, tuning the radio frontend, and delivering the first batch of processed PSD data.

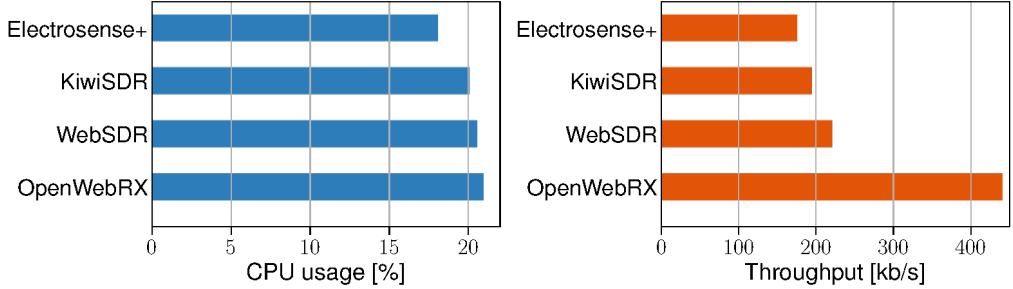

Figure 5: Comparison between Electrosense+ and other existing projects. CPU usage (left) and network throughput (right) of the client while PSD and AM decoder are enabled.

We have also evaluated the response time for the decoders, meaning that we measure the average delay between the moment a user selects a decoder until the first decoded payload arrives to the web-browser. Every time a new decoder is selected by the user, the sensor requires to stop the previous decoder, re-tune to a new frequency, and start the new decoder (among other tasks). For radio FM/AM decoders, the average waiting time until the user starts receiving and listening to the audio stream is 2.6 seconds. Once the FM/AM decoder is set, the retuning by the user is much faster since there is no need to start the decoder again. For the rest of the decoders that stream a data flow encoded as JSON (e.g. ADS-B), the average waiting time for the user is 1.5 seconds.

### 5.3. Automatic Channel Identification

Using the PSD data sent by the sensor, the web-based client infers where active channels are located in the spectrum by applying a power-based channel identification algorithm. The idea of this feature is to support the user to identify spectrum interesting parts which are utilized and can potentially be decoded. We have evaluated the automatic channel identification algorithm performance over FM radio bands. As Table 4 shows, using a 5 seconds averaging window over the PSD stream data, our model is 84% accurate with a precision of 0.88 (low false positive rate). Applying an averaging window over the PSD data has two positive effects: 1) reducing the noise and thus improving the algorithm performance, and 2) requiring less computation power on the client side.

Table 4: Automatic channel identification performance (FM Radio)

|  | TP | TN | FP | FN | Accuracy | Precision | Recall | F1 |
| --- | --- | --- | --- | --- | --- | --- | --- | --- |
| no-avg | 14 | 31 | 2 | 9 | 0.80 | 0.88 | 0.61 | 0.72 |
| avg-5sec | 14 | 33 | 2 | 7 | 0.84 | 0.88 | 0.67 | 0.76 |

*5.4. Rewarding model*

We perform an evaluation of the rewarding model detailed in Section 3.3. We have also added an illustrative example for a better understanding. Figure 6 (top) shows the owner earnings, depending on the activated pipeline, with respect to the density of sensors, time the sensor was deployed and operational time. The earnings are normalized with respect to the maximum number of tokens available at any time. In the study, the network establishes the maximum operating time for the decoding pipeline in each sensor to 80% of the total operational time.

**Network earnings, costs and benefits.** The network must be self sustainable making sure that generates enough benefits in order to deal with the network cost. Figure 6 (bottom) shows the network earnings (users spending tokens to access the decoding pipeline), network costs (sensor owners receiving tokens by keeping their sensors operative), and network benefits. The Electrosense+ network has a profit margin applying this reward model as long as the decoding pipeline is enabled 23% of the time in average. This value can be tuned by increasing or decreasing the benefits of the network.

**Illustrative example.** Let us assume that the network was created 500 days ago, therefore $T = 500$. The rewarding model is evaluated every week taking in to account the sensor density, deployment time and operational time. For the sake of simplicity, the network is composed by only 2 sensors located in different cities. Both sensors were deployed at $t_s = 1$ (when the network started). For sensor1 the operational time of the PSD pipeline is 7 days, and 1 day for the decoding pipeline. For sensor2 the operational time of the PSD pipeline is 7 days, but in this case it has 3 operational days in the decoding pipeline. The earning of the network is obtained by charging tokens (K) for decoding pipeline consuming. Every week, the rewarding system is evaluated to compute earnings of both network and users. In our example, during the previous week 4 days of decoding operation were observed among all sensors of the network. Assuming that $P_{dec} = 5K$, $P_{psd} = 1K$, $F = 0.4$. In common, the two sensors have been operating in decoding for 4 days, and in PSD for 10 days. The network have charged $4 days * P_{dec} * (1 + F) = 28\ K$ to other users for using decoding pipeline. The earnings of the network are $F * 28\ K = 11.2\ K$, therefore we have 16.8 K to paid the sensor owners rewards for this week. The operational time of decoding is paid 5 times higher that the operational time of psd, therefore $P_{dec} = 16.8 * (5/6) = 14\ K$, $P_{psd} = 16.8 * (1/6) = 2.8\ K$. The difference between the two sensors of this toy example is the operational time on the decoding pipeline. Therefore the system will reward with more tokens to sensor2 due to the higher operational time in the decoding pipeline. Sensor1 will be rewarded by 1.4 tokens (PSD) and 3.78 tokens (decoding), while sensor2 will be rewarded by 1.4 tokens (PSD) and 8.82 tokens (decoding).

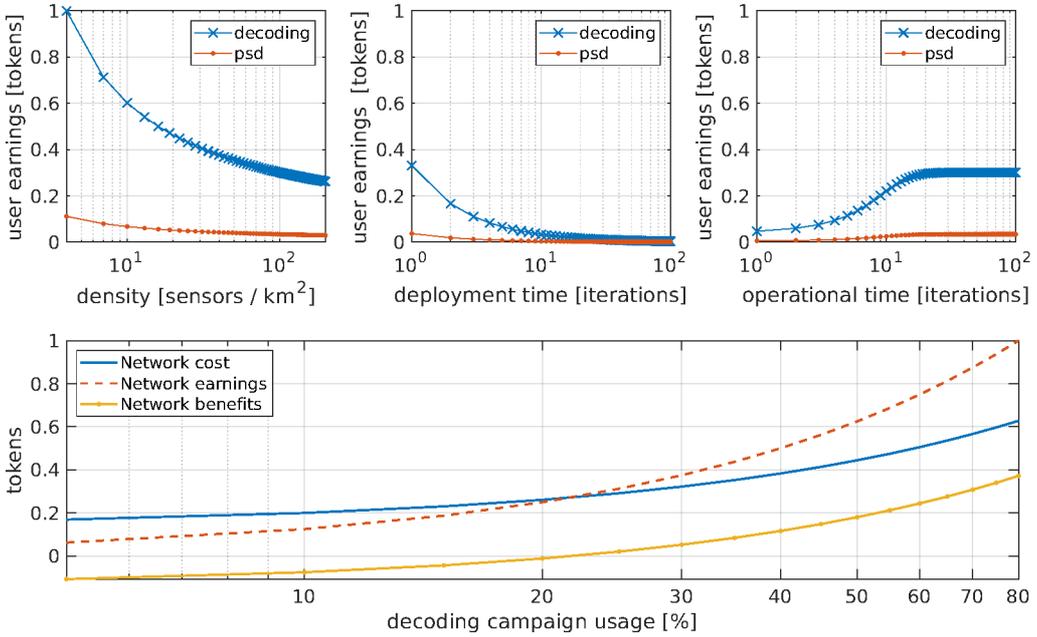

Figure 6: Rewarding system for sensor owners (top) and network costs, earnings and benefits (bottom). Earnings are normalized to the maximum of tokens available.

*5.5. Scalability*

Electrosense+ architecture takes advantage of peer-to-peer communication between the sensor and the user. The system scalability depends on the traffic load of the signaling server (see Figure 1) which handles the control messages (connection request, keep alive, etc). These messages represent less than 2 kb/s per sensor, meaning that one instance of the signaling server with a 50 Mb/s symmetric network can manage more than 25K sensors at once.

From the sensor point of view, the current version of Electrosense+ supports one single user connected to the sensor at the same time to avoid the saturation of the network connection of the sensor owner. To avoid abuse for one single user we have implemented the reward model that limits somehow the time that a single user can be connected to a sensor. In addition, multiple user connections asking for different decoders are not allow since Electrosense+ sensors have only one radio front-end.

# 6. Conclusion

We have presented Electrosense+, a system that allows users to remotely decode specific parts of the radio spectrum using IoT sensors. Electrosense+ is built on top of its centralized-approach predecessor [1], and provides a new peer-to-peer communication among clients and sensors to exchange information and make the system scalable. We have proposed a reward model to provide incentives to people to deploy and keep sensors running while making the network deployment sustainable over time. The system architecture allows to decode any broadcast wireless signal that is within range of the sensors. We have integrated several publicly available decoders that are not intrusive to the privacy of the wireless users. The decoders operate on the sensor-side and we have optimized their computational performance to run in embedded and IoT devices. We manage to keep the average CPU load of the IoT sensors below 40% in most of the cases, even when the PSD and decoding pipeline are executed on the sensor at the same time. The communication channel is also implemented in an efficient way, which allows to keep the network bandwidth low between the sensor and the client: for streaming a single audio channel to the user the bandwidth needed is 50 kb/s, while for sending data (e.g. generated by the ADS-B decoder) the bandwidth used can go up to 200 kb/s. In both cases the network bandwidth is low, allowing the users to connect to the system using conventional home Internet connections and WiFi hotspots. We have implemented a friendly web-interface (platform independent) to easily interact with the sensors. Electrosense+ provides the opportunity to gain better knowledge and understanding of the spectrum utilization, by offering remotely signal decoding capabilities and direct incentives for sensor owners to deploy spectrum sensors and keep them running.

## Acknowledgments

This research work by IMDEA Networks Institute was sponsored in part by armasuisse under the Cyber and Information-Research-Program, the NATO Science for Peace and Security Programme under grant G5461, and Madrid Regional Government through TAPIR-CM project S2018/TCS-4496.